\documentclass[prl,twocolumn, showpacs, preprintnumbers,amsmath,amssymb]{revtex4}

\usepackage{graphicx}
\usepackage{hyperref}

\begin{document} 

\title{Charge, density and electron temperature in a molecular ultracold plasma}
\author{C. J. Rennick, J. P. Morrison, J. Ortega-Arroyo, P. J. Godin, N. Saquet and E. R.  
Grant}
\affiliation{Department of Chemistry, University of
British Columbia, Vancouver, BC V6T 1Z1, Canada}

\date{\today}

\begin{abstract}
A Rydberg gas of NO entrained in a supersonic molecular beam releases electrons as it evolves to form an ultracold plasma.  The size of this signal, compared with that extracted by the subsequent application of a pulsed electric field, determines the absolute magnitude of the plasma charge.  This information, combined with the number density of ions, supports a simple thermochemical model that explains the evolution of the plasma to an ultracold electron temperature.  
\end{abstract}


\maketitle

Under suitable conditions, the quasi-neutral system of ions and electrons that forms an ultracold plasma can develop the properties of a strongly coupled fluid.  This intriguing degree of charged-particle correlation occurs when the average interparticle Coulomb energy exceeds the thermal energy, or $\Gamma > 1$, where $\Gamma$~ quantifies the ratio of Coulomb to thermal energy in the ion and electron subsystem reservoirs:  $\Gamma_{i,e} =(q^2/4\pi\epsilon_0 a)\slash kT_{i,e}$, in which the Wigner-Seitz radius, $a$, relates to the particle density by, $\frac{4}{3}\pi a^3 = 1/\rho$.  

For $\Gamma \approx 1$, a plasma can form liquid-like correlations \cite{Ichimaru}.  Beyond a correlation parameter of about 180, simulated one-component plasmas undergo a phase transition to form body-centered-cubic crystals \cite{Dubin}.  The creation of an experimental plasma with properties such as these would provide an important laboratory analog for the strongly coupled systems that appear under conditions of extreme temperature and density, such as in quark-gluon plasmas, at the centres of heavy stars and in the plasmas generated by the interaction of high-power lasers with solid targets.  

Most often, ultracold plasma experiments start with a magneto-optical trap (MOT) in which an atomic gas sample with an initial temperature lower than $100 ~\mu K$ absorbs light from a high-resolution laser to release electrons with an average energy of $10~K$ or less \cite{Killian_Phys_rept}.  In the plasma that forms, three-body electron-ion recombination drives the initial electron temperature to a value above the strongly coupled regime, typically $T_{e}>30 ~K$.  

In a novel approach, we have excited nitric oxide cooled in a seeded supersonic beam to form a durable $molecular$ ultracold plasma with an ion density exceeding $1\times 10^{12}$~cm$^{-3}$ \cite{Morrison2008}, and an electron temperature, estimated from a Vlasov analysis, to fall with expansion from an initial value as low as $7~K$ to $1 ~K$ or less over a lifetime that extends beyond $100~ \mu$s \cite{Morrison2009}.  

Such conditions of temperature and density suggest an electron coupling parameter substantially greater than $1$.  If confirmed, this degree of correlation would represent a significant milestone in ultracold plasma research.  The molecular beam method offers a versatile means to study plasmas built both on atomic and a great many molecular precursors, and the well defined propagation of the plasma volume in the laboratory frame presents significant advantages for imaging.  

The present letter reports a measurement of the absolute excess charge of this plasma, and uses this information to construct a simple thermochemical model for its evolution from a Rydberg gas.  This model serves to confirm our estimate of the total charged-particle density and suggests a mechanism by which this system attains such a low electron temperature.  Calculations relate this feature to the high plasma densities achievable in a molecular beam.  

\begin{figure}[htb]
	\begin{center}
		\includegraphics[width=\columnwidth]{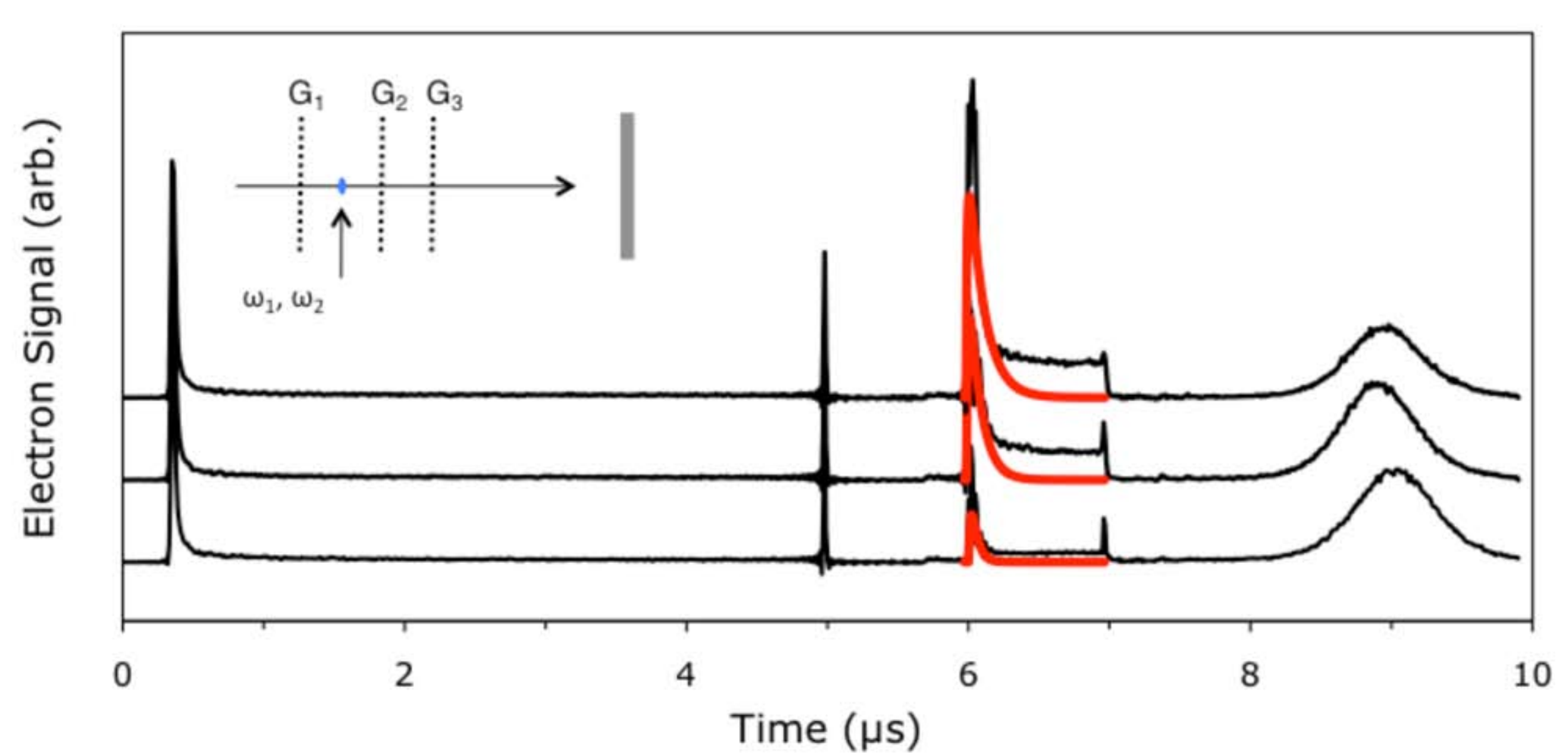}
	\end{center}
	\caption{(Color online) Ultracold plasma electron signal waveform.  Prompt pulse, reverse-bias switching transient, pulsed field signal (from bottom:  37.5, 50 and 100 V cm$^{-1}$) and plasma electrons extracted on transmission through G$_2$.   Simulations model pulsed-field electron signal waveforms, neglecting the baseline elevation observed at higher extraction voltages.  Inset:  Molecular beam path through grids to a microchannel plate detector.}
	\label{fig:scope}
\end{figure}

Our experimental apparatus expands nitric oxide, seeded at 10\% in helium, through a 0.5~mm diameter nozzle.   A 1 mm diameter skimmer selects the core of this jet to form a molecular beam in the second stage of a differentially pumped electron spectrometer.  Two frequency-doubled pulsed dye lasers intersect this beam 8 cm beyond the skimmer, between a pair of field grids, labeled G$_{1}$ and G$_{2}$ in Fig. \ref{fig:scope}. The first laser excites NO from its ground state to the $v = 0$, $N=0$ level of the A($^2\Sigma^+$) state. From this gateway state, the second dye laser drives the system of molecules to the $n=51$, $l=3$ Rydberg state converging to the $N^+ = 2$ rotational state of the ion ($51f(2)$), which lies well below the lowest $v^+ = 0$, $N^+=0$ ionization limit. At sufficient density, these Rydberg molecules interact, leading to a sequence of ionization processes that form an ultracold plasma \cite{Morrison2008}.  

This laser-crossed molecular beam illumination geometry creates a 750 $\mu$m by c.a. 2 mm prolate ellipsoid excitation volume.  The cross-beam density falls off rapidly along the long axis of this ellipsoid, and for convenience, we approximate the initial plasma density distribution as a spherical Gaussian with a radius, $r_0=375~ \mu$m.  Assuming saturated steps of double-resonant excitation, the density of NO at this point in the beam yields a molecular Rydberg density as high as 5$\times$10$^{12}$ cm$^{-3}$, or a total of approximately 10$^9$ Rydberg-state NO molecules in the initial excitation volume.  

We observe plasma formation heralded by a small signal of prompt electrons.  A bias applied to G$_{1}$ maintains a very weak field (-250 mV cm$^{-1}$) between G$_{1}$ and G$_{2}$, which extracts these promptly released electrons for detection.  Subsequently, the plasma moves with the velocity of the molecular beam toward the detector to reach G$_{2}$, which is grounded.   On passing through this plane, the G$_{2}$ grid extracts electrons, which accelerate in the 100  V cm$^{-1}$ field between G$_{2}$ and G$_{3}$ to reach the detector.  The resulting waveform measures the electron density profile of the plasma along its axis of propagation.

A voltage pulse applied between G$_{1}$ and G$_{2}$, before the plasma reaches G$_{2}$, serves as a probe to gauge the absolute magnitude of its excess positive charge.  Using a Behlke high-voltage switch, we elevate G$_{1}$ to a predetermined negative voltage for an interval of 1~$\mu$s, 6~$\mu$s after the lasers have fired, which sends a pulse of extracted electrons to the detector.  The grids return to near ground before the plasma passes through G$_{2}$.  For best reproducibility, we sometimes find it useful to precede this pulsed-field extraction with a 5 V cm$^{-1}$ reverse-bias pulse to sweep away electrons weakly bound to the low-density wings of the plasma volume.  Fig. \ref{fig:scope}  shows a waveform displaying the prompt signal of electrons associated with plasma formation, the electrons extracted from the plasma by the pulsed field applied as it travels between G$_{1}$ and G$_{2}$, and the electrons shed as the plasma traverses the G$_{2}$ grid.  

The prompt signal represents the $Q_0$ electrons released by energy pooling collisions, including Rydberg-Rydberg Penning ionization, and subsequent electron-impact ionization of Rydberg molecules.  Electrons liberated by these processes escape the illuminated volume until the potential energy of the accumulating excess positive charge, $N^{*}\sqrt{2/\pi}~q_{e}^{2}/4\pi\epsilon_{0}r_0$ exceeds $\frac{3}{2}~k_{B}T'_e$, where $T'_e$ represents the temperature of these initially formed energetic electrons and $N^*$ is the number of positive ions required to reach the trapping potential threshold for a plasma of radius $r_0$ \cite{Killian1999,Comparat_star}.  

Electron-impact ionization continues in an avalanche \cite{Pohl2003} that cools trapped electrons \cite{Pohl2006} until this process exhausts their energy or the population of Rydberg molecules.  Electrons can also gain energy by three-body electron-cation recombination and superelastic collisions that de-excite Rydberg molecules, but dissociative channels, open for molecules, may diminish the importance of these latter effects.  

We see from the initial signal that prompt loss depletes the plasma of a certain number of electrons, producing a net positive space charge.  It remains now to determine the absolute magnitude of this charge, and reconcile its appearance with the electron temperature evidenced by the rate of plasma expansion \cite{Morrison2009}.  

By the application of pulsed field, we both tip the long-range potential and polarize the plasma, adding a dipole component to the angular dependence of its potential at short range.  A full description of these effects would require a self-consistent solution of the Poisson equation for the field-distorted electron density distribution, with the resulting calculated potential sensitive to the selected ion distribution \cite{King_model}.   We neglect this complexity, and instead, for computational as well as conceptual simplicity, we model the potential outside the plasma as that of a charged, conductive sphere in an external field:

\begin{equation}
	V(r, \theta) = \frac{Q}{4\pi\epsilon_0 r} - E_0 r \cos\theta + E_0 
	\frac{r_{0}^3}{r^2} \cos\theta\,\mathrm{,}
	\label{equ:potential}
\end{equation}

\noindent where $r_{0}$ is the radius of the sphere, $Q$ is the total charge and $E_0$ is the externally-applied field.  This approximation in effect: i) assumes that the high charge density of the plasma core screens its interior electrons, and ii) collapses all the electron density beyond a critical radius to form a surface charge distribution from which the pulsed field extracts electrons.  

The plasma charge combines with the external field to create a saddle point that defines $\epsilon_\mathrm{crit}(\theta)$, the potential energy an electron on the $r_0$ surface must overcome to escape.  We calculate this flux of pulsed-field extraction electrons in steps by considering the trapping potential for each element of surface area, and then integrating over area and electron kinetic energy for an assumed thermal distribution at temperature, $T_e$.  For each step, we recalculate the charge and raise the lower limit, $\epsilon_\mathrm{crit}(\theta)$, in the integral over electron kinetic energy for each element of surface area.  

Elastic collisions between the trapped electrons and ions transfer energy inefficiently owing to their large mass difference, so the ions remain at a low temperature. However, electron-electron thermalization proceeds on the subnanosecond timescale of a few plasma periods \cite{Robicheaux2002}. We therefore assume that the electron kinetic energy distribution function is always Maxwellian.  Thermalizing collisions fill-in the truncation that occurs as high kinetic-energy electrons from the tail of the distribution leave the plasma.  

\begin{figure}[htb]
	\begin{center}
		\includegraphics[width=\columnwidth]{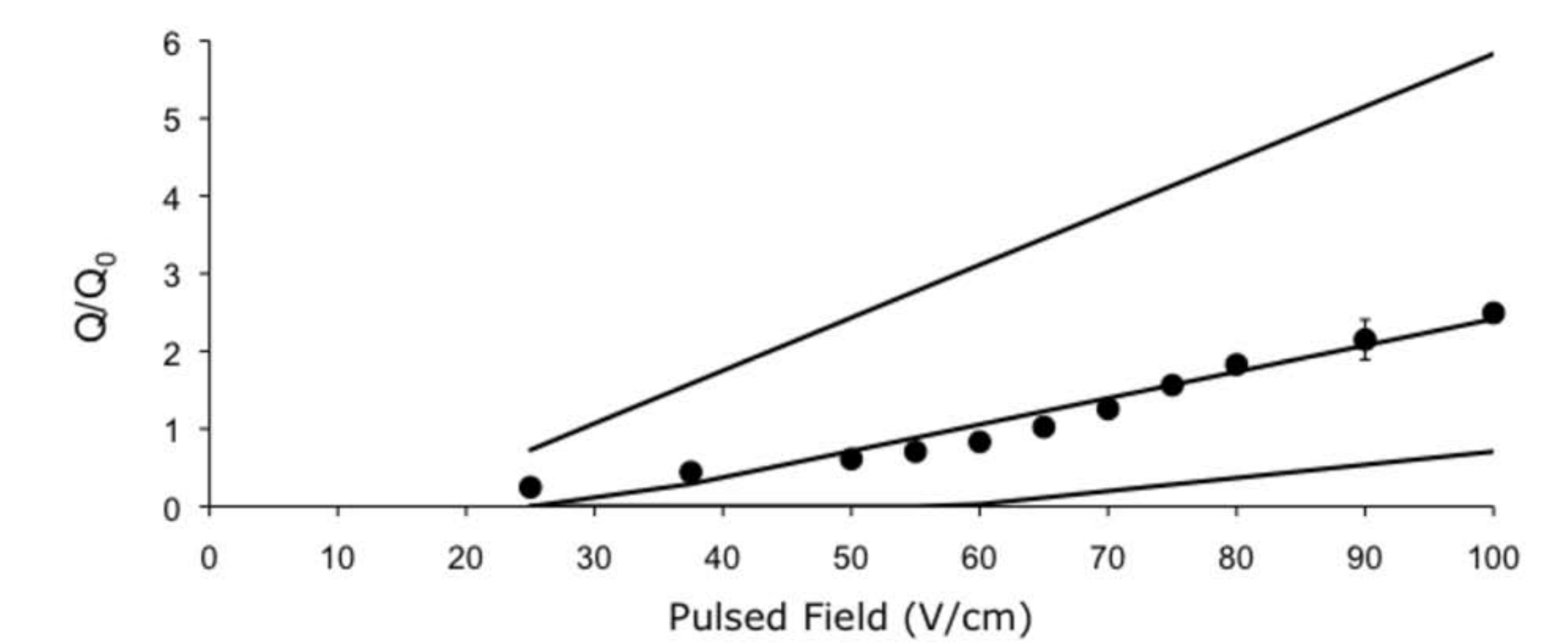}
	\end{center}
	\caption{Signal versus pulsed-field normalized by the plasma charge.  Measured data (solid circles) normalized by the prompt signal.  Calculated lines normalized by $Q_0$ chosen for the simulation. Line through the data points for an initial charge, $N_i - N_e$, of $Q_{0}=9.5\times 10^5$.  Upper and lower lines assume $Q_{0}= 4.75\times 10^5$ and $1.9\times 10^6$ respectively.}
	\label{fig:calibrated}
\end{figure}

To simulate the experimental data, we select an initial plasma charge, $Q_0$ and calculate the total number of electrons, $Q$, extracted at each applied field.  If the $Q_0$ we choose is too small, this calculation overestimates the pulsed field signal relative to $Q_0$.  If we choose too large a $Q_0$, the predicted $Q/Q_0$ is too small.   Figure \ref{fig:calibrated} compares experimental ratios of pulsed field to prompt signals with simulations scaled in this way for three values of $Q_0$.  A least-squares fit of the simulation to the experimental data yields an estimated value of $Q_0$ corresponding to the initial loss of $9.5\times10^{5}$ electrons.

A sequence of two events determines this excess charge and the observable electron temperature of the plasma.  Penning collisions first convert pairs of NO Rydberg molecules into a free NO$^+$ ion, an electron and a de-excited NO molecule, which very likely predissociates to form translationally excited neutral atoms.  The excess energy of ionization creates an initial population of electrons with an average energy characterized by some temperature $T'_e$.  This energy enables electrons to escape until the attractive potential of the developing NO$^+$ space charge traps the electrons that remain.  A cascade of collisions then excite and de-excite remaining NO Rydberg molecules, giving rise to an avalanche of electron-impact ionization events.  Because, on balance, these electron liberating collisions consume energy, the trapped electrons cool and remain confined \cite{Vanhaecke, Pohl2006}.  

A few assumptions enable us to develop a simple model for this evolution to a plasma.  Without loss of generality, we can regard the stages of Penning and electron-impact ionization to occur in sequence or overlap in time.  The initial escape of Penning electrons evaporatively lowers $T'_e$, but we assume the loss of comparatively few confines the rest, making this effect negligible.  During avalanche ionization, exothermic electron-Rydberg deactivating collisions can occur, but this energy pooling process simply raises the phenomenological temperature we ascribe to the Penning electrons.  Unlike atoms, NO molecules in Rydberg states cascading to lower principal quantum number rapidly predissociate, and we assume that product N and O atoms leave the interaction volume without affecting the electron or ion temperatures.  

Experimentally, Killian et al.\cite{Killian1999} found that they could describe the fraction of electrons confined by various plasmas of defined excess energy as a uniform function of the number of ions $N'_i$, scaled by the number, $N^*$, required at threshold in each case to trap electrons.  This function, confirmed by simulations \cite{Comparat_star}, relates $N^*$ to $N'_i - N'_e$, the excess charge or number of electrons lost, and $N'_i$, the number of ions (as produced in the current instance by Penning ionization):
\begin{equation}
	N^* = \frac{(N'_i - N'_e)^2}{N'_i}\,\mathrm{.}
	\label{equ:NStar}
\end{equation}
As defined above, the threshold number $N^*$ increases for plasmas of higher electron temperature and larger $r_0$ radius as, 
\begin{equation}
	N^* = \frac{3}{2}k_BT'_{e}r_0\frac{4\pi\epsilon_0}{q^2_e}\sqrt{\frac{\pi}{2}}\,\mathrm{.}
	\label{equ:NStarT}
\end{equation}

Our experiment measures the final excess charge, $N_i - N_e$, which we equate to the early time excess charge, $N'_i - N'_e$, on the assumption that no avalanche electrons escape.  We can accurately estimate the total number of Rydberg molecules, produced at saturation, $N_0$, from the properties of our molecular beam.  We have previously measured the electron temperature of the evolved plasma, $T_e$ by determining its expansion rate \cite{Morrison2009}.  

Starting with $N_i - N_e$ as currently measured, together with known $N_0$ and $T_e$, we can examine the energy balance for various trial values of $f_P$, the fraction of Rydberg excited NO molecules consumed by Penning ionization.  Each trial value of $f_P$ sets a  balance between $N'_i$, the number of initially formed ions, calculated as, $N'{_i}=f_{P}N_{0}/2$, and $N''_{i}$, the number of excited NO molecules remaining to be ionized by the electron-impact avalanche, as
\begin{equation}
	N''_{i} = N''_{e}=(1-f_{P})N_0\mathrm{.}
	\label{equ:N''_i}
\end{equation}
\noindent Having $N'_i$, we obtain $N^*$ from Eq. \ref{equ:NStar}.  We can then use Eq. \ref{equ:NStarT} ~to determine a temperature $T'_e$ of the Penning electrons.  

Each ion produced by electron impact ionization of a nitric oxide molecule in the $51f(2)$ Rydberg state consumes the binding energy  $\Delta E_n = 8.38\times10^{-22}$ J.  Thus, avalanche ionization to form $N''_i$ ions lowers the electron temperature of the plasma to a value, $T_e$, such that:

\begin{equation}
	T_e = \frac{\frac{3}{2}k_BT'_eN'_e -  N''_e\Delta E_n}{\frac{3}{2}k_B(N'_e + N''_e)}\,\mathrm{.}
	\label{equ:Te}
\end{equation}

$N'_{e}$ approximately equates to $N'_{i}$ over a wide range of $f_P$, and we can use Eqs.  \ref{equ:NStar} -- \ref{equ:N''_i} to write a simple expression for $T_e$ in terms of the measured charge, $N_i - N_e$, and the fraction, $f_P$, of Rydberg NO molecules consumed by Penning ionization.  

\begin{equation}
	T_e = \frac{q_e^2(N_i - N_e)^{2}/\sqrt{2\pi}r_{0}2\pi\epsilon_0-(1-f_{P})N_0\Delta E_n}{\frac{3}{2}k_BN_0(1-f_{P}/2)}\,\mathrm{.}
	\label{equ:Tef}
\end{equation}

\begin{figure}[htb]
	\begin{center}
		\includegraphics[width=\columnwidth]{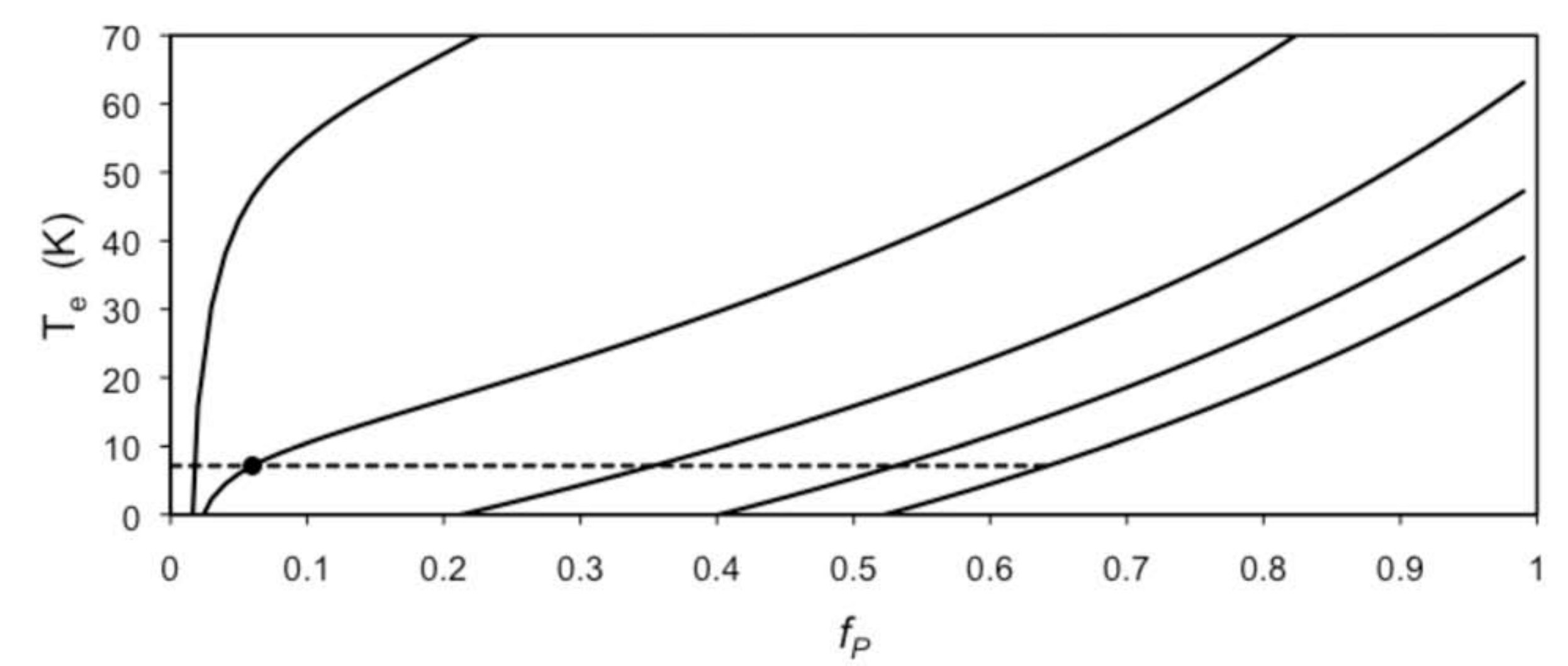}
	\end{center}
	\caption{Electron temperature upon evolution to plasma predicted from thermochemical energy balance as a function of $f_P$, for assumed values of initial Rydberg density, $N_0$, reading from the left, of 1, 2, 3, 4 and 5 $\times 10^{12}$ cm$^{-3}$.  Below $f_{p}=0.05$, Eq. \ref{equ:Tef}~ necessarily incorporates additional terms to account for the difference between $N'_{i}$ and $N'_{e}$.}
	\label{fig:Tvsf}
\end{figure}

Fig. \ref{fig:Tvsf} shows values of $T_e$ predicted by Eq. \ref{equ:Tef} as a function of $f_P$ for assumed values of $N_0$, decreasing from a saturated $N_0$ = $1.1\times 10^9$ (density, $n_e=5\times10^{12}$ cm$^{-1}$).  

This variation in plasma electron temperature reflects the simple thermochemical balance in which electron heating, manifested in the observed excess charge, dissipates in the cooling necessary to complete avalanche ionization.  For the purposes of illustration, consider an excitation volume containing a population of NO molecules excited to the $51f(2)$ Rydberg state with a density of $2\times10^{12}$ cm$^{-3}$.  Under these circumstances the Wigner-Seitz radius is 500 nm, compared with an orbital radius of 130 nm.  The corresponding Erlang distribution function of nearest-neighbour distances predicts that slightly more than 6 percent of this population will exist as pairs spaced by 150 nm or less \cite{Chotia}.  It seems reasonable to assume that this population of proximal molecules will interact to produce Penning electrons on a very short timescale, and that these electrons will both in part escape to create a trapping potential and initiate an endothermic electron avalanche.   

Thus, for $N_0 = 4.4\times10^{8}$, an assumed $f_{P}=0.06$ yields $N'_{i}=1.3\times10^{7}$, which produces the measured excess charge, $(N_i - N_e)=9.5\times10^{5}$ for a $T'_{e}$=1600 K.  Electron-impact ionization of the Rydberg molecules that remain consume $3.5\times10^{-13}$ J of electron kinetic energy, lowering $T_{e}$ to 7 K.  A point plotted on Fig. \ref{fig:Tvsf} indicates this $N_0, f_P$ intersection.  The selected value of $N_0$ reflects either a smaller than saturated initial preparation or just the number of NO molecules ionized when Rydberg predissociation overtakes ionization and cooling stops.  

This simple model neglects some obvious aspects of the plasma evolution dynamics.  For example, as mentioned above it considers neither the reduction in $T'_e$ that occurs as a consequence of early electron escape, nor increase in $T_e$ by electron-impact deactivation of NO Rydberg states.  However, we know that the plasma loses comparatively few hot electrons to form its trapping potential, and we can assume that collisional deactivation of Rydberg NO accelerates predissociation, which rapidly depletes the plasma of molecules in quantum states below the bottleneck for collisional ionization \cite{Mansbach, Kuzmin}.

Regardless of its simplifying assumptions, this crude energy balance does serve to show that a feasible route exists for the molecular plasma of NO$^+$ ions and electrons formed in a molecular beam to reach a low electron temperature.  The variation in the cooling range with $N_0$ shows clearly how high density plays a critical role in providing a sufficient reservoir of Rydberg bound states to cool the system as it evolves.  Fortunately, typical conditions attained in pulsed dye laser crossed differentially pumped molecular beams yield excited state densities in this range for many molecular systems.  

The authors gratefully acknowledge helpful discussions with Thomas Pohl.  This work was supported by NSERC, the Canada Foundation for Innovation (CFI) and the BC Knowledge Development Fund (BCKDF).

\bibliography{pulsedField}

\enddocument